\documentclass[a4paper]{jpconf}
\usepackage{graphicx}
\usepackage{amsmath}
\usepackage{hyperref}
\begin{document}
\title{Calculating entropy by Ma's method for a system of $k$-mers on a square lattice}

\author{Denis V Kokosinskii$^1$, Mikhail V Ulyanov$^{1,2}$}
\address{$^1$ Computational Mathematics and Cybernetics, M~V~Lomonosov Moscow State University, Moscow 119991, Russia}
\address{$^2$ V~A~Trapeznikov Institute of Control Sciences of RAS, Moscow 117997, Russia}

\ead{deniskokossskappa@gmail.com}
\ead{muljanov@mail.ru}

\begin{abstract}
Random walk of rectangular particles on a square lattice leads to a pattern formation when the hard-core interaction between the particles is assumed. To estimate changes in the entropy during this random walk, we propose a modification of Ma's method.  
The 2D sliding window technique was used to divide a system under consideration into subsystems. We used Ma's ``coincidence'' method to estimate the total number of possible states for such subsystems. In this study, the accuracy of Ma's method is studied in a simple combinatory model, both experimentally and theoretically. We determine which definition of ``coincidence'' for this scheme leads to greater accuracy. 
Ma's estimate of the number of possible states for a system of $k$-mers correlates well with the estimate obtained using a ``naive'' method.
\end{abstract}

\section{Introduction}\label{sec:intro}
Boltzman's entropy $S$ is defined as
$S = k_B \ln(N),$
where $N$ is the number of possible microstates, corresponding to the system's macrostate, and $k_B$ is the Boltzmann's constant. Boltzmann's entropy is an important characteristic of any evaluating system. Calculating Boltzmann's entropy directly as the logarithm of the total number of microstates for a current macrostate is difficult for large systems. For complex, dense systems, direct calculation of the Boltamzann's entropy may require a great deal of computation.  The obvious solution to this problem is to obtain only an estimate for the entropy, as this is faster to calculate. Therefore, approximations of entropy, that are easier to compute, can help. Different ways to estimate entropy were proposed~\cite{Ma1981,Roma2001,Ulyanov2018,Vogel2020}.

In our study, we propose a modification of Ma's method~\cite{Ma1981}, apply it to a system of $k$-mers, and investigate its accuracy. In~\sref{sec:modelandmethod}, the system under consideration is described, the modification of Ma's method, and the accuracy of Ma's method is investigated in a simple combinatory model. In~\sref{sec:application}, we apply Ma's method to a system of $k$-mers. \Sref{sec:concl} summarizes the main results.

\section{Model and Methods}\label{sec:modelandmethod}
\subsection{System under consideration}\label{subsec:system}

A square lattice of size $256$ by $256$ cells is used as a substrate.  At moment 0, elongated rectangular particles ($k$-mers) of size $1 \times k$ or $k \times 1$ lattice cells are placed onto the substrate using random sequential adsorption (RSA)~\cite{Evans1993}. The RSA process~\cite{Evans1993} stops when no further $k$-mer can be added to the lattice. A Monte Carlo simulation of the random walk of $k$-mers is then begun. We shall, here give only a brief description of this process, although a full description can be found at~\cite{Lebovka2017}. One Monte Carlo step is $M$ attempts to shift a randomly selected $k$-mer in a random direction by 1 lattice cell, where $M$ is the total number of $k$-mers on the lattice. A $k$-mer can shift either along its' orientation or perpendicular to it, rotation being forbidden. The $k$-mers are ``blind'', they are not aware of the positions of other $k$-mers. If the position after the shift would cover a position already occupied by other $k$-mers, the shift is not made and the shift is attempted with a different $k$-mer. To avoid boundary effects, the lattice is considered to have toroidal boundary conditions. Thus, a $k$-mer moving beyond the right boundary of the lattice will appear on the left.

\begin{figure}[!htb]
\centering
\includegraphics[width=15cm]{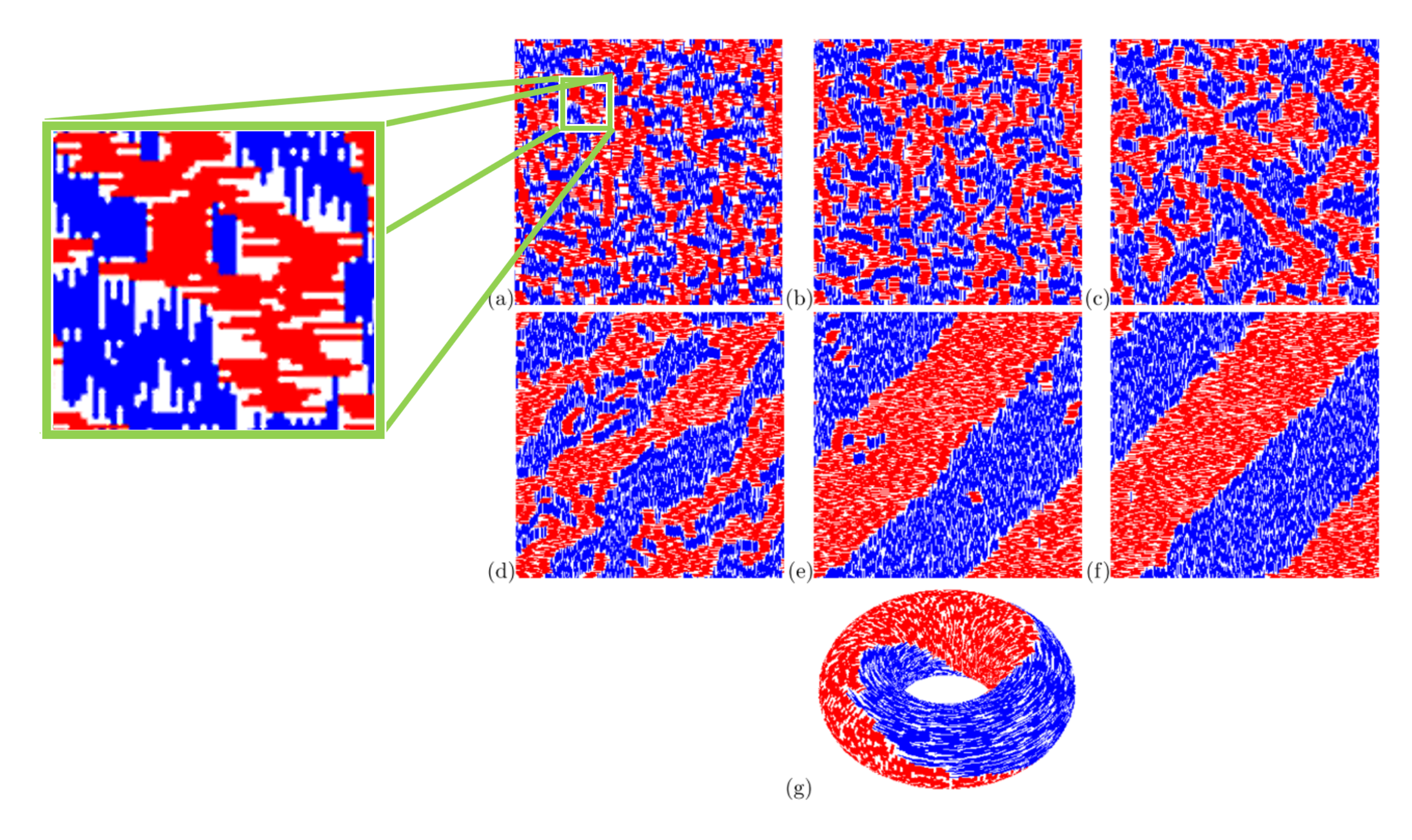}
\caption{Temporal evolution of the system. (a) $t_{MC} = 0$, initial jammed state, (b) $t_{MC} = 100$, end of fluidization, (c) $t_{MC} = 10^4$, labyrinth
patterns, (d) $t_{MC} = 10^5$, labyrinth patterns are transforming into stripes, (e) $t_{MC} = 10^6$, final stage of stripe formation, (f) $t_{MC} = 10^7$, steady state in the form of
diagonal stripes, (g) final ``yin–yang'' pattern on a torus (the same pattern as in (f)); for clarity, the major radius of the torus is significantly exaggerated. Illustration adopted from~\cite{Tarasevich2017}. }
\label{fig:system}
\end{figure}

The evolution of the system of $k$-mers on the lattice is illustrated in~\fref{fig:system}. Seemingly, this isolated system evolves from chaos to order. Random walk of rectangular particles of two mutually perpendicular orientations on a square lattice with periodic boundary conditions demonstrated a pattern formation. The system tends to self-organize, as shown in~\cite{Tarasevich2017}. This pattern formation is supposed to be entropy-driven. Since the complexity of counting the possible microstates grows exponentially with the number of $k$-mers in the system, any direct computation of entropy changes is hardly reasonable. An estimation of the entropy seems promising.

The complexity of the direct calculation of the Boltzmann's entropy of this system depends exponentially on the total number of $k$-mers. A ``naive'' estimate of entropy at any given step can be calculated as
$$
S = \sum_p \ln(\text{DoF}(p)),
$$
where $p$ is any $k$-mer, and $\text{DoF}(p)$ is the number of degrees of freedom of the particle $p$, in other words, the number of directions, in which $p$ can shift. Boltzmann's constant is omitted here and subsequently, because we are interested in the relative values of entropy at the different steps. This estimate does not take into consideration potential blocking of $k$-mers shifts after some previous $k$-mers have moved.

\subsection{Ma's method in a simple combinatory model}\label{subsec:method}
Ma proposed estimating the total number of system states by counting the number of ``coincidences'', i.e. paired microstates. We aim to investigate this method in a simple combinatory model.
Let $N$ be a total number of boxes, $n$ be the number of balls thrown at these $N$ boxes with uniform distribution, and $N_p$ be the number of resulting pairs (coincidences) in boxes. The index of a box where the ball is thrown can be interpreted as this ball's microstate, so the total number of registered microstates would equal $n$. $N$ would then be the total number of possible microstates of the system. Suppose $n$ and $N_p$ are given and we would like to determine $N$. According to Ma, $N$ can be estimated as
\begin{equation}
\label{eq:ma}
	N = \frac {n(n-1)}{2N_p}.
\end{equation}
A question arises of how to correctly count the pairs. We can count pairs as the number of boxes with exactly two balls ($N_{p1}$) or as the number of boxes with two or more balls ($N_{p2}$). Another approach is to count all the possible pairs in each box and then sum these values for all boxes ($N_{p3}$). If $k$ balls are in a box, there are $\binom{k}{2}$ possible pairs of them.  It is important to notice, that in the case of $n \ll N$, all three approaches produce close results. The probability of exactly $k$ balls in any box can be calculated as
\begin{equation}
\label{eq:pk}
	p_k = \binom{n}{k} {\left(\frac {1}{N} \right)}^k {\left(1 - \frac {1}{N}\right)}^{n-k} ,
\end{equation}
so
$$
	\frac {p_3} {p_2} = \frac {n-2}{3N} \left(1 - \frac {1}{N}\right).
$$
In the case $n \ll N$, $p_3 \ll p_2$. 

\begin{figure}[!htb]
\begin{minipage}[c]{0.5\textwidth}
  \centering
\includegraphics[width=\textwidth]{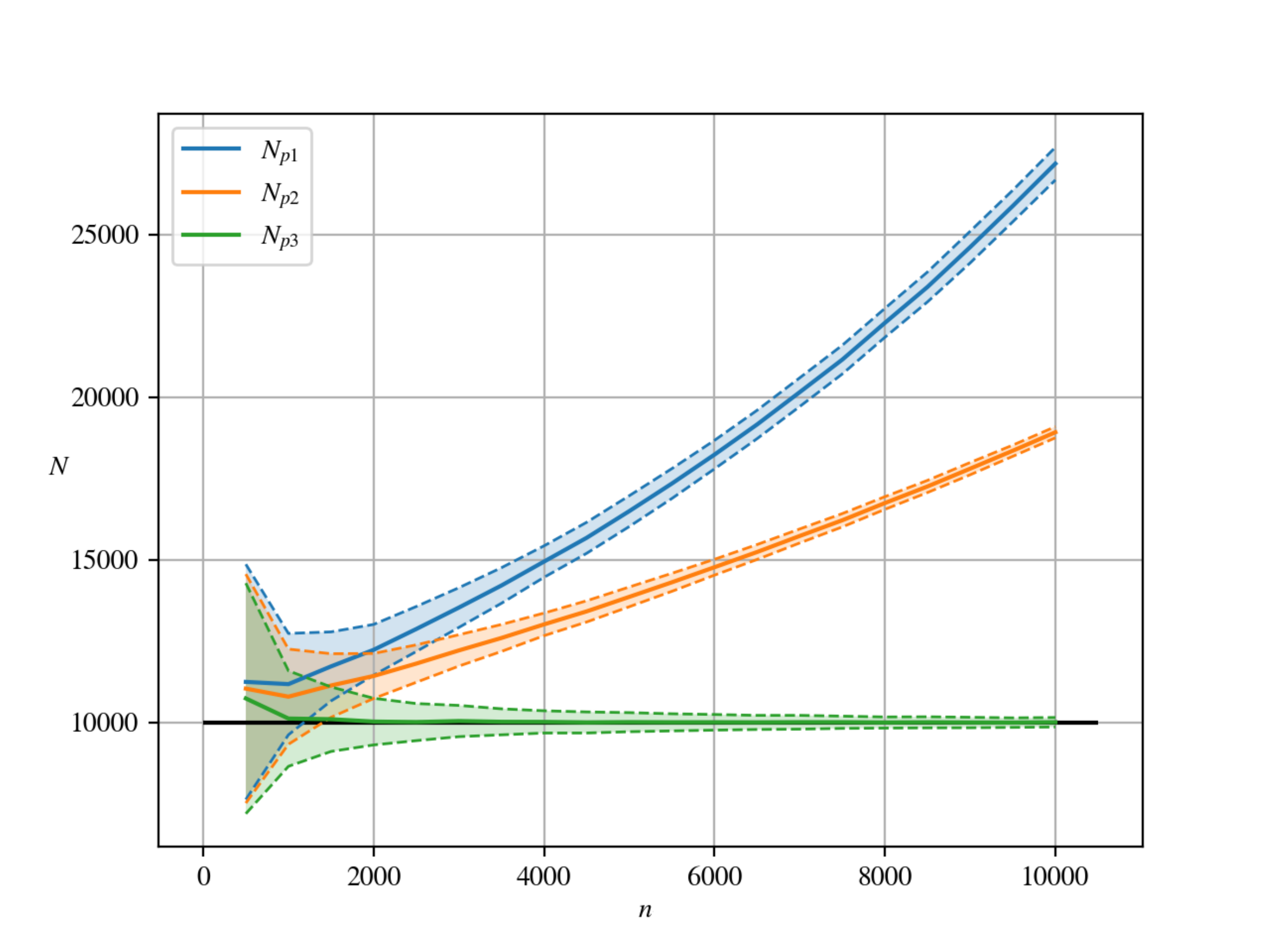}\\
\end{minipage}
\hfill
\begin{minipage}[c]{0.45\textwidth}
\caption{$N$ estimation using equation~\eref{eq:ma} with different approaches to pair counting. Number of boxes $N=10 000$. Estimate averaged over $1 000$ experiments for each value of $n$.}
\label{fig:paircounts}
\end{minipage}
\end{figure}

The results of experiments for estimating $N$ using different pair counting strategies are presented in~\fref{fig:paircounts}. It is noticeable that method 3 produces good estimation of $N$ even when $n \sim N$. The errors of the other methods appear to grow as $n$ increases. Let us examine this method in terms of probability theory.
Let $Z_i$ be a random variable, $Z_i = \binom{k}{2}$, when $k$ balls are in a given box $i$. Using~\eref{eq:pk}, we can calculate the mathematical expectation of $Z_i$:
\begin{equation}
\label{eq:EZi}
	E(Z_i) =  \sum_{j=2}^{N} {\binom {j}{2} p_j}.
\end{equation}
We can rewrite $p_k$ as
$$
	p_k = \binom{n}{k} {\left(\frac {1}{N} \right)}^k {\left(1 - \frac {1}{N} \right)}^{n-k} = \left[ \, \frac {\binom {n}{k} }{ \binom {n}{2}} {\left(\frac {1}{N}\right)}^{k-2} {\left(1 - \frac {1}{N}\right)}^{-k+2} \right] \, \cdot p_2
$$
or
\begin{equation}
\label{eq:pkfromp2}
	p_k = \left[ \, \frac {\binom {n}{k} }{ \binom {n}{2} } {\left( \frac {1}{N-1} \right)}^{k-2} \right] \, p_2.
\end{equation}
Substituting $p_k$ in \eref{eq:EZi} with~\eref{eq:pkfromp2} we get
$$
	E(Z_i) = \sum_{j=2}^{N} {\left[ \, \binom {j}{2}  \frac {\binom {n}{j} }{ \binom {n}{2} } {\left(  \frac {1}{N-1} \right)}^{j-2} p_2 \right] \,}.
$$
We then use the properties of binomial coefficients' and move the common factor out
$$
	E(Z_i) = p_2  \sum_{j=2}^{N} {  \left[ \, { \binom {n-2}{j-2} } {(\frac {1}{N-1})}^{j-2} \right] \, }
$$
and finally apply the binomial formula
\begin{equation}
\label {eq:EZiFinal}
	 E(Z_i) = p_2 {\left( 1 + \frac {1}{N-1} \right)}^{n-2}.
\end{equation}
Now, if we introduce a random variable $Z=\sum_{i=1}^{N} {Z_i}$, it takes the values of $N_{p3}$. The mathematical expectation of $Z$ is
$$
	E(Z) = \sum_{i=1}^{N} {E(Z_i)} =  N p_2 { \left( \frac {N}{N-1} \right)}^{n-2}.
$$
Using representation~\eref{eq:pk}, we get:
$$
	E(Z) = N \left[ \, \binom {n}{2} {\left( \frac {1}{N} \right)}^2 {\left( \frac {N - 1}{N} \right)}^{n-2} \right] \, {\left (\frac {N}{N-1} \right)}^{n-2} = \binom {n}{2} \frac {1}{N} = \frac {n(n-1)}{2N} .
$$
We can now get an estimate of N:
$$
	N = \frac {n(n-1)}{2 E(Z)}
$$
We have proved that the formula $N=\frac {n(n-1)}{2 N_{p3}}$ provides an unbiased estimate of $N$ regardless of $n$.

\section {Results: Application of Ma's method to a system of $k$-mers}\label{sec:application}
To apply Ma's method, we first need to define a microstate for the system of $k$-mers on the lattice. The lattice can be subdivided into small non-overlapping squares. We can define a microstate as the system state inside one of these squares. However, in this case, the results of computer experiments show, that no pairs (equal states of the system in different squares) can be found at any step of a Monte Carlo simulation of random walk. The existence of at least one pair is required to apply Ma's method.

\begin{figure}[!htb]
\begin{minipage}[c]{0.6\textwidth}
  \centering
\includegraphics[width=\textwidth]{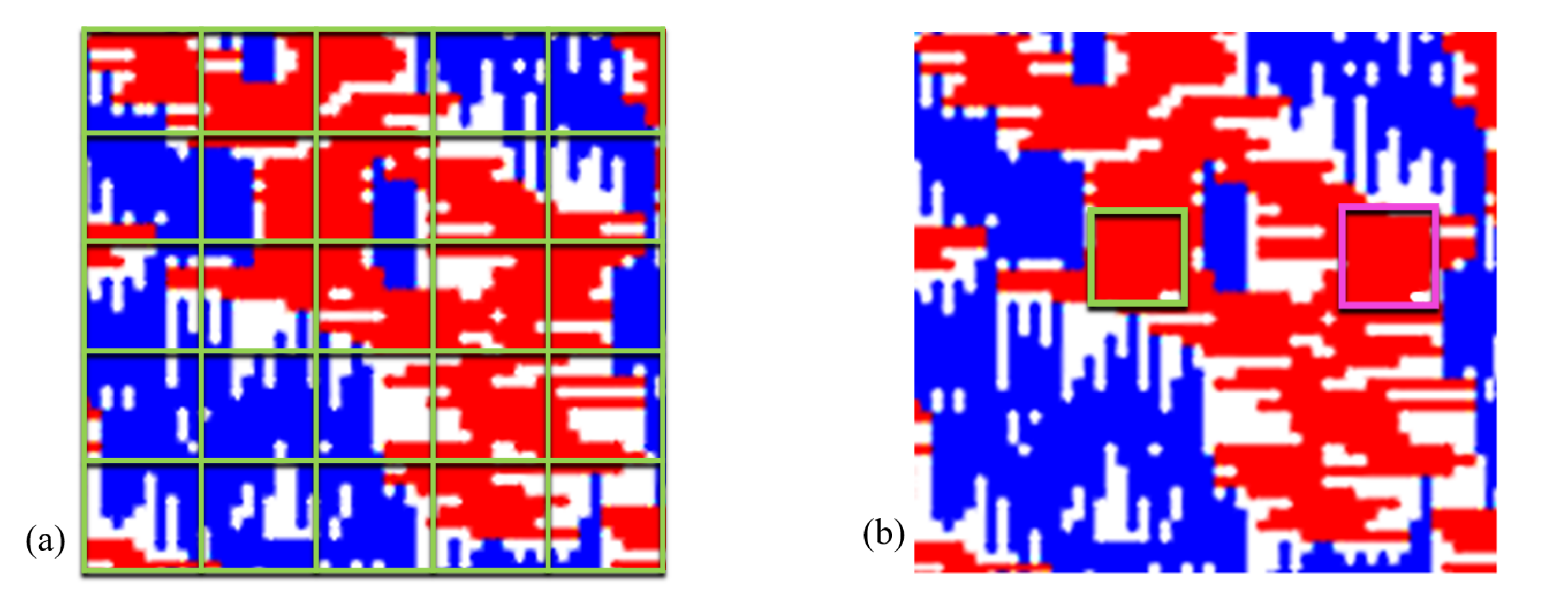}\\
\end{minipage}
\hfill
\begin{minipage}[c]{0.35\textwidth}
\caption{Two definitions of a microstate for a system of $k$-mers. (a) non-overlapping squares; (b) 2D sliding window. A ``paired'' microstate is shown in (b). }
\label{fig:windowdef}
\end{minipage}
\end{figure}

We can use another definition of a microstate, based on~\cite{Ulyanov2018}. The 2D sliding window approach was used in this article to investigate the various structural properties of the same system and process that we are studying. At any given Monte Carlo step, a small square window of size $L$ by $L$ lattice cells moves over all the positions on the system. There is a total of $256 \times 256$ positions of the window, because the boundary conditions are toroidal. We can count identical states of the window in different positions and use Ma's method to estimate the total number of possible window states. Here, and subsequently, we count the number of pairs according to the best pair counting approach from~\sref{sec:modelandmethod}. The two definitions of a microstate are illustrated in~\fref{fig:windowdef}.

\begin{figure}[!htb]
\centering
\includegraphics[width=0.9\textwidth]{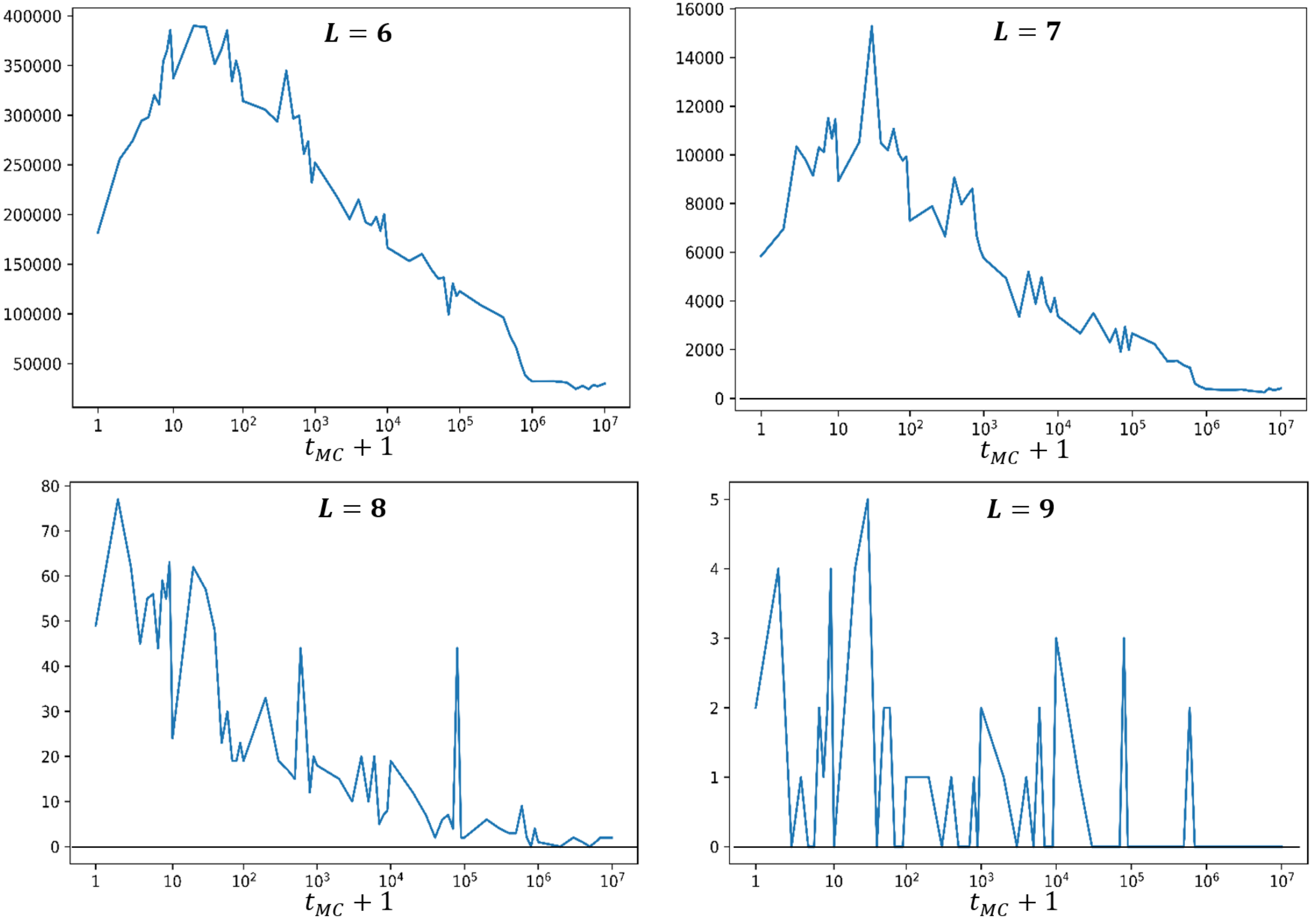}
\caption{The total number of pairs for different window sizes $L$ during the Monte Carlo simulation of random walk for the system of $k$-mers.}
\label{fig:windowsize}
\end{figure}

We need to choose the window size $L$. There are two important considerations for $L$. First, there must be at least one pair of window states at any Monte Carlo step. Second, the bigger the size of the window, the better it represents the state of the whole system. The total number of pairs of window states during the Monte Carlo process for various window sizes $L$ is presented in~\fref{fig:windowsize}. A window size $L=7$ satisfies the requirements.

\section{Conclusion}\label{sec:concl}
\begin{figure}[!hbt]
\begin{minipage}[c]{0.5\textwidth}
\centering
\includegraphics[width=\textwidth]{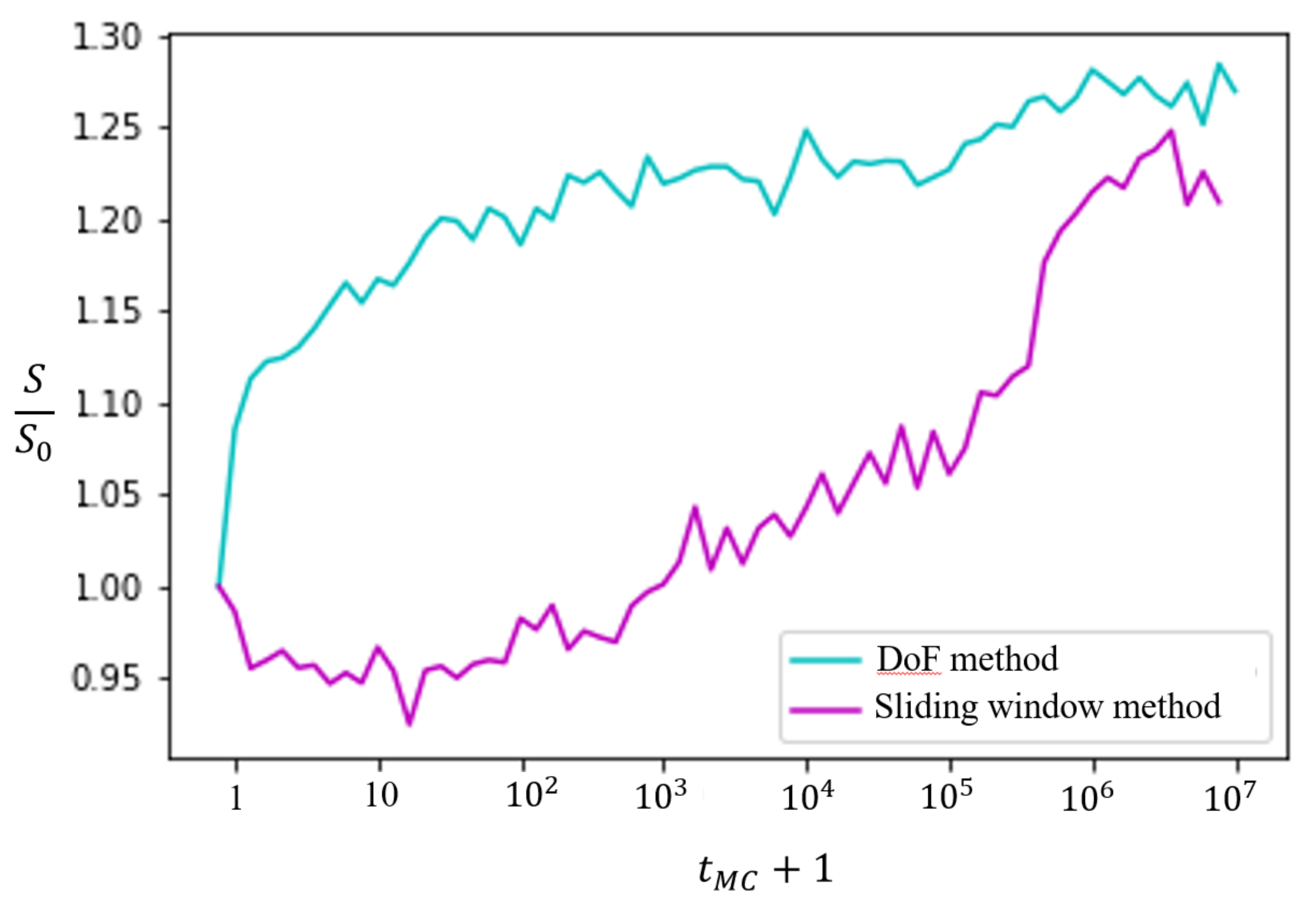}\\
\end{minipage}
\hfill
\begin{minipage}[c]{0.45\textwidth}
\caption{Entropy estimates during the Monte Carlo simulation of the diffusion of $k$-mers on a lattice. Entropy is normalized to the entropy at step 0.}
\label{fig:results}
\end{minipage}
\end{figure}

Our study demonstrates, that, in the system under consideration, the rearrangement of particles and pattern-formation are entropy-driven, i.e., in its initial state, a particle has less possibilities to change its location in comparison with any succeeding state~\cite{Frenkel1993}. The Ma's entropy estimate for a 2D sliding window during a Monte Carlo simulation of diffusion of a system of $k$-mers on a lattice is presented in~\fref{fig:results}. The ``naive'' method, introduced in~\sref{sec:intro}, is also presented as the baseline. Both estimates grow, despite the fact, that the system self-organizes. Worthy of note, that Ma's estimate is, in practice, several magnitudes smaller, than the ``naive'' estimate. As the positions of the sliding window overlap, so it may not be correct to assume, that its' states are independent and uniformly distributed. However, Ma's method, if using the correct definition of a pair from~\sref{sec:modelandmethod}, will give close estimates if these conditions are satisfied. It might be worth testing to estimate the entropy in systems where the microstates are independent and uniformly distributed.

\ack
We acknowledge funding from the Russian Foundation for Basic Research, Project No.~18-07-00343. We are thankful to Yuri Tarasevich for his stimulating discussions.

\section*{References}
\providecommand{\newblock}{}

\end{document}